\documentstyle[12pt,aasms4,amssym]{article}

\newcommand{\etal}{et al.}
\newcommand{\gx}{\mbox{GX\hspace{0.5ex}1+4}}
\newcommand{\ang}{$\stackrel{\circ}{\rm A}$}

\lefthead{Kotani et al.}
\righthead{ASCA and Ginga Observations of GX 1+4}

\begin{document}

\title{ASCA and Ginga Observations of \gx}

\author{Taro Kotani}
\affil{kotani@riken.go.jp\\
Cosmic Radiation Laboratory, RIKEN, Wako, Saitama, 351-0198, Japan}
\author{Tadayasu Dotani, Fumiaki Nagase}
\affil{dotani@astro.isas.ac.jp, nagase@astro.isas.ac.jp\\
Institute of Space and Astronautical Science, Sagamihara, 
Kanagawa, 229-8510, Japan}
\author{John G. Greenhill}
\affil{John.Greenhill@utas.edu.au\\
Physics Department, University of Tasmania, GPO Box 252C, Hobart,
Tasmania, Australia}
\author{Steven H. Pravdo}
\affil{Steven.H.Pravdo@jpl.nasa.gov\\
Jet Propulsion Laboratory, 4800 Oak Grove Drive, M/S: 306-438,
Pasadena, CA 91109-8099, USA}
\and
\author{Lorella Angelini}
\affil{angelini@lheavx.gsfc.nasa.gov\\
Laboratory for High Energy Astrophysics, NASA/GSFC Code 668,
Greenbelt MD 20771, USA}

\begin{abstract}
The X-ray binary pulsar \gx\ was observed with Ginga every
year from 1987 to 1991, and with ASCA in 1994. During the Ginga
observations, \gx\ was in the steady spindown phase, though the
X-ray flux was not steady. Assuming a distance of 10 kpc, the
absorption-corrected X-ray luminosity decreased down to $L_{\mbox{2-20
keV}} = 2.7 \times 10^{36}$ erg s$^{-1}$ in 1991, after the peak
activity of $L_{\mbox{2-20 keV}} = 1.2 \times 10^{37}$ erg s$^{-1}$ 
in 1989. On the other hand, the absorption column density
showed a drastic increase over the Ginga observation series. It
was less than $10^{23}$ cm$^{-2}$ at the beginning of the series, and
it reached a maximum of $(1.4 \pm 0.2) \times 10^{24}$ cm$^{-2}$ in
1991, indicating a rapid accumulation of matter in the vicinity of the
source. The center energy and equivalent width of the iron line
were consistent with emission by isotropically distributed cold matter. The ASCA
observation was performed on  1994 September 15, a month before the
transition into a spinup phase. The source brightened again to
$L_{\mbox{2-20 keV}} = 10^{37}$ erg s$^{-1}$. The absorption column density
was observed to decrease for the first time to $(2.08 \pm 0.02)
\times 10^{23}$ cm$^{-2}$. The ionization degree of iron in the
absorbing matter was determined to be Fe {\sc i}--Fe {\sc iv} using
the ratio of the line-center energy to the absorption-edge energy. The
low ionization degree is consistent with an absorbing matter
distribution extending $\sim 10^{12}$ cm from the source.  We compared
the results with optical observations, and found the optical data also
supports the picture.  Based on the geometrical model, possible causes
of the bimodal behavior of the source are discussed.

\end{abstract}


\section{Introduction}
The X-ray pulsar \gx\ is the most enigmatic of all binary X-ray pulsars.
It is identified with the symbiotic system V2116 Oph, which contains
a magnetized neutron star accreting from the M6III companion
(Davidsen, Malina \& Bowyer 1977).
\gx\ is a slow pulsar with a spin period of about 2~minutes.
It has a hard energy spectrum, and hence has been a good target for both
soft and hard X-ray observations.

In the 1970's, \gx\ was a bright X-ray source with a typical 2--6 keV
flux of 100 mCrab (McClintock \& Leventhal 1989 and references therein).
(Hereafter, the bandpass of flux measurement is 2--6 keV, if not specified.)
It showed the fastest spinup among the binary X-ray pulsars with 
a mean rate of $\dot{P}/P = -2.7 \times 10^{-2}$ yr$^{-1}$
(Elsner \etal\ 1985).
In the early 1980's, the source entered into a low state.
EXOSAT observations in 1983 failed to detect the source and set
an upper limit of 0.5 mCrab (McClintock \& Leventhal 1989 and references therein).
Ginga detected \gx\ in 1987 at a flux level of 2~mCrab and
the source was found to be spinning down (Makishima \etal\ 1988).
Subsequent Ginga and HEXE observations showed that the spindown rate was
remarkably constant at $\dot{P}/P = 1.3 \times 10^{-2}$ yr$^{-1}$ 
in spite of a gradual increase  by an order of
magnitude in the X-ray flux (Sakao \etal\ 1990; Mony \etal\ 1991).
The clear contrast of \gx\ in the 1970's and in the 1980's indicates
the close connection  between the high/low transition in luminosity and the 
spin reversal.

The spindown rate was observed to decrease a little in 1990 and 1991,
although subsequently a linear trend of spindown was recovered with a
slightly increased rate (Mandrou \etal\ 1994).  A large flare in the hard
X-ray band was detected by CGRO in 1993 September (Staubert \etal\ 1995).and 
the source brightened up to $\sim$80~mCrab (40--100~keV) for a few
weeks. A gradual increase of the spindown rate had been observed  for about 6 months before
the flare (Finger \etal\ 1993), but the rate was
constant during the flare at $\dot{P}/P = 3.2 \times 10^{-2}$
yr$^{-1}$ (Staubert \etal\ 1995).

The spin period history of \gx\ has been almost continuously monitored by
BATSE since 1991 (Chakrabarty \etal\ 1997).  It revealed a short episode
of spinup in 1994 (Chakrabarty et al.\ 1994), which was accompanied by
a large increase in the hard X-ray flux.

The 20--100 keV phase-averaged pulsed flux reached $\sim$100~mCrab in 1994 November.
The spinup phase was terminated in 1995 March (Chakrabarty \etal\ 1995)
and the spindown resumed at a rate of $\dot{P}/P = 4.8 \times 10^{-3}$ 
yr$^{-1}$.

The origin of this enigmatic behavior in \gx\ is not yet understood.
Even the basic parameters of the system, such as the orbital period
and the surface magnetic field of the neutron star, are not known.  A
binary period of 304~day was suggested by Cutler \etal\ (1986) based
on the fluctuation of the spinup rate.  However, optical monitoring of the 
H$_{\alpha}$ emission line indicates that the period may be
significantly greater than 304~days (Sood \etal\ 1995).  If the spin
behavior is analyzed in the framework of the accretion torque model by
Ghosh and Lamb (1979a, 1979b) and Wang (1987), a very large magnetic
field on the neutron star surface, $\sim\!\!10^{14}$~G, is required
(Makishima \etal\ 1988).  Dotani \etal\ (1989) showed that even a
moderate magnetic field of a few $\times10^{12}$~G can explain the
spindown rate, if a retrograde accretion disk were formed around the
neutron star.  An energy-dependent pulse profile in hard X-rays is
used to construct a model based on two-photon emission as the dominant
continuum source (Greenhill \etal\ 1993).  The retrograde disk model
is supported by the BATSE observations (Nelson \etal\ 1997), although
some other possibilities such as the transition between Keplerian flow
and advective flow in accretion geometry are also suggested (Yi,
Wheeler \& Vishniac 1997).

In this paper we report results obtained from the unpublished Ginga 
and new ASCA data.

\section{Observations}
The data we analyzed in this paper were obtained with Ginga (Makino \&
the Astro-C Team 1987) and ASCA (Tanaka \etal\ 1994).  The main
instruments aboard Ginga are the Large Area Counters (LAC; Turner \etal\
1989), which are non-imaging, conventional proportional counters. The LAC has an effective area of 4000~cm$^2$ and 
covers an energy range of 1--37~keV or 2--60~keV depending on the high
voltage setting It has the
relatively poor energy resolution of $\sim$18~\% at 6~keV\@.  The field of
view of the LAC is restricted by collimators to $1^{\circ} \times
2^{\circ}$ (FWHM)\@.

Ginga observations of \gx\ were made on 5 occasions.  The first 3 sets
of data were already analyzed and the results were previously
published (Makishima \etal\ 1988; Dotani \etal\ 1989; Sakao \etal\
1990).  We used the 4th and 5th sets of data for the present analysis.
The fourth observations were carried out on 1990 September 4--5, as a
part of galactic ridge observations, and the 5th ones were on 1991
September 16.  In the 5th observation, the high voltage setting of the
LAC was changed to cover 2--60~keV, while in the 4th observation the
nominal setting of 1--37 keV was adopted.  Both sets of data were
acquired in MPC1 mode with 48 energy channels.  In this data mode the 
temporal resolution was 0.5, 4, and 16~s for high, medium, and low bit
rate, respectively.  Unfortunately, most  of the 5th
observation was recorded in low bit rate mode, and this limited the
temporal resolution in later analysis.

ASCA carries four identical grazing-incidence X-ray telescopes (XRT),
and each XRT has an imaging spectrometer at its focal plane.  The XRT
utilize nested thin-foil conical reflectors to achieve large effective
area (1200~cm$^2$ at 1~keV; Serlemitsos \etal\ 1995).  A point source
image has a typical size of $3'$ in half power diameter.  The focal
plane detectors are two solid-state spectrometers (SIS; Burke \etal\
1994) and two gas imaging spectrometers (GIS; Makishima \etal\ 1995).
The SIS consists of 4 CCD chips optimized for X-ray photon
detection.  It covers an energy range of 0.5--10~keV and a single chip
covers a $11' \times 11'$ sky field.  The SIS achieves high energy resolution of 2~\% (FWHM) at
6~keV\@.  The GIS is a gas scintillation proportional counter with
imaging capability.  It covers an energy range of 0.7--10~keV and the
field of view of $25'$ in radius.  The GIS has an energy resolution of
7.8~\% (FWHM) and a spatial resolution of 0.5~mm (FWHM) both at
6~keV\@.

ASCA observations of \gx\ were carried out in 1994 September 14--15.
The observation mode of SIS was 1-CCD faint mode, and that of GIS was
PH mode with standard bit assignment.  Thus the time
resolution of SIS was 4 s and that of GIS was 62.5 ms and 0.5 s for
high and medium bit rate, respectively.  The journal of the
observations are summarized in Table~1.

\placetable{kotani:tbl:obslog}

\section{Analysis and Results}
\subsection{Ginga data}
Raw data were first screened according to the good time intervals,
which exclude South Atlantic Anomaly passages, Earth occultation of
the source, and the periods of low cut-off rigidity ($<$6 GeV/c).
Light curves of the data were then checked to remove parts of the data
contaminated by particle events manually.  These screened data were
used for the subsequent analysis.

\subsubsection{Timing analysis}
The pulse period of \gx\ was determined using an epoch folding technique.
We did not subtract background from the data, because the  timescale
of the background variation was much longer than $10^2$~s and did not
affect the period determination of \gx. The pulse period which gives
the largest $\chi^2$ against a constant was searched. The time resolution of
the data was 4~s in the 1990 observation, so we divided the folding
pulse period into 28 phase bins. The time resolution was 16~s in
the 1991 observation and we divided the folding period into 7 phase
bins. Finer binnings were not appropriate because the pulse period is
about $10^2$ s.  Heliocentric pulse periods we obtained are listed in
Table~2. Average rates of pulse period change between the previous
measurement and this measurement are also listed in Table~2. These results
indicate that \gx\ was spinning down since 1987.

\placetable{kotani:tbl:tim}

The folded light curves are shown in Figure~1.
The pulse profile of data set (a) is very similar to those in 1988 and 1989:
double peaked in low energy band and single peaked in high energy band
accompanied by a sharp and deep dip.
The sharp dip observed in data set (a) is not clear in data set (b).
This may be due to the low time resolution of the data.

\placefigure{kotani:fig:lc}

\subsubsection{Spectral analysis}
\gx\ is located close to the galactic plane, hence it is difficult
to accurately estimate the contribution from the galactic ridge
emission (Koyama 1989 and the references therein).  For this
estimation, we sampled the spectra at two points in the sky, ($l$,
$b$) = ($3^{\circ}.215$, $5^{\circ}.744$) and ($5^{\circ}.795$,
$7^{\circ}.674$); \gx, ($1^{\circ}.943$, $4^{\circ}.785$), is located
on the line connecting these two points.  The data were obtained as a
part of galactic ridge observations, in which the data of \gx\ were
also acquired.  These two points are well separated from nearby bright
sources such as the Kepler SNR and GX~9+9, and hence the spectra can
be regarded as those of the pure galactic ridge emission. The spectra
were fitted by a model of thermal Bremsstrahlung and a Gaussian line
with a center energy of 6.7~keV - the energy of Fe {\sc xxv}
K$\alpha$.  The temperature of the thermal Bremsstrahlung was found to be
4.78 keV and 2.14 keV, respectively, and the flux in unit of count
rate at spectral peak was $2.5\pm0.7$ counts s$^{-1}$ keV$^{-1}$ and
0.95 counts s$^{-1}$ keV$^{-1}$, respectively.  The Gaussian line was
$0.89\pm0.31$ counts s$^{-1}$ at the former position and not detected
at the latter.  The flux of the galactic ridge emission at the
position of
\gx\ was then estimated by linear interpolation of the best-fit
parameters.  The resultant temperature and spectral peak flux were
6.1~keV and 3.27 counts s$^{-1}$ keV$^{-1}$, which corresponds to
1--37 keV flux of 18.1 counts s$^{-1}$.  The iron line flux was
estimated to be 1.2 counts s$^{-1}$.  The ridge emission thus
determined was subtracted from the energy spectra of \gx\ together
with the intrinsic and Cosmic X-ray background utilizing the method
described in Hayashida \etal\ (1989). The resultant,
background-subtracted spectra are shown in Figure~2.

\placefigure{kotani:fig:specginga}

The energy spectrum of the observation in 1991 was well fitted by a
power law attenuated by a cold absorber, as was the case in the three
previous observations in 1987, 1988, and 1989. The spectrum of 1990
was, however, better expressed by a partial covering model of a power
law continuum.  A prominent iron line was also detected at about 6.4
keV\@.  This line is considered to be a fluorescence line of neutral
or slightly ionized iron in the vicinity of the source, not a
collisionally excited emission line of highly ionized iron whose
center energy is at 6.7 or 6.9 keV\@.  The fluorescence line provides
important information on the physics of the system (Makishima 1986).
The center energy gives the ionization stage of the illuminated matter
and thus the distance of the plasma from the source and the density.
The geometry of the attenuating matter can be derived from the
equivalent width of the line and the attenuation column density.
Therefore, it is essential to resolve a 6.4 keV line from 6.7 and 6.9
keV lines, and determine the central energy and equivalent width
precisely.   Unfortunately the 1991 energy spectrum has coarse energy binning 
and the line center energy could not be determined with enough
accuracy to distinguish the 6.4 keV and 6.7 keV lines.  As seen later,
ASCA data show that the iron line in \gx\ is centered at 6.4 keV\@.
Thus, in the subsequent analysis of Ginga spectra, we fixed the line
center energy to 6.4 keV\@.  We summarize in Table~3 the best-fit
parameters of both sets of observations. The spectra and the best-fit
model are shown in figure~2.  We detected significant X-ray emission
from \gx\ up to $\sim$50~keV in data set~2. However, we could not find
any significant structure of cyclotron resonant scattering in the
energy spectrum.

\placetable{kotani:tbl:best}

The absorption-corrected X-ray luminosity is calculated 
as $L_{\mbox{2-20 keV}} = 0.72 \times 10^{37}$ erg s$^{-1}$ and $0.27
\times 10^{37}$ erg s$^{-1}$ for 1990 and 1991, respectively, for the
assumed distance of 10 kpc. 
In 1989, the source was found to be as luminous as $10^{37}$ erg s$^{-1}$.
It was interpreted that \gx\ was recovering from the low state (Sakao
\etal\ 1990). 
In 1991, however, it decreased again to the very faint
level as in 1987, and was still in the steady spindown phase.

\subsection{ASCA data}
\gx\ was detected with both GIS and SIS with an average count rate
of about 1 counts s$^{-1}$ (GIS: 1--10~keV).  No significant time
variation was seen except for the 122~s pulsation.  Source photons were
extracted from the circular region centered on the GIS image peak with a
radius of 6~arcmin.  As for SIS, photons from the whole chips
(S0C1 and S1C3) were used to get better statistics. Both timing and
spectral analyses were done using these source photons.

\subsubsection{Timing analysis}
The pulse period of \gx\ was determined using the GIS data, which have
superior time resolution.  We applied neither detector-background
subtraction nor dead time correction to the data, because such
corrections are not significant for timing analysis of medium
brightness sources such as
\gx.  Using the epoch folding technique, we found the barycentric
pulse period to be $122.019\pm0.027$~s.  Folded pulse profiles are
shown in figure~3 in several different energy bands. Deep dips similar
to that observed with Ginga are clearly seen in the pulse
profile. Below 3~keV, galactic ridge emission dominates and hence the
pulse amplitude is much reduced.  It should be noted that the dip
structure in the pulse profile becomes shallow in the energy band
between 6.28--6.52 keV, where the fluorescent iron line is dominant
(see Figure~4).  Iron line flux contributes about 40 \% in this energy
band and the reduction of the dip depth ($\sim$50 \% in 6.28--6.52 keV
compared to $\sim$80 \% in the other bands) is consistent with no
pulsation in the iron emission line.  This conforms to the scenario
that the fluorescent iron line originates from the spherically
symmetric absorbing matter as discussed in section~4.  The other
energy bands besides $<$3~keV and 6.28--6.52 keV show a similar pulse
profile and pulse fraction, indicating no pulse phase dependence of
the spectral parameters, such as the photon index or the absorption
column density.

\placefigure{kotani:fig:lcasca}

\subsubsection{Spectral analysis}
We calculated a phase-averaged energy spectra of SIS and GIS using the
source photons described previously. For the GIS spectrum, background
photons were extracted using the GIS data from the region symmetric to
the source region relative to the bore sight axis of the XRT\@. These
background data include galactic ridge emission. We constructed an
energy spectrum of the galactic ridge emission by subtracting the high
latitude background data, which are publically available. We found that
the background spectrum can be well represented by  emission
from an optically thin hot plasma model by Mewe \etal\ (1985). The best-fit
temperature obtained was 6.1~keV with equivalent hydrogen column
density of $2.3\times10^{21}$ cm$^{-2}$.  Because we could not extract
background data for SIS due to the small FOV, we just subtracted the
high latitude blank sky data as background. Galactic ridge emission
was included in a model spectrum for spectral fitting.

We found that a single power law modified by cold matter absorption
plus a narrow emission line can reproduce the energy spectrum.
The model functions we employed are
\begin{equation}
f_{\rm SIS}(E) = f_{\rm Mewe}(kT) \exp(-\sigma N_{\rm H}^{\rm IS})  
+ \left [ A \;E^{-\Gamma} + B\exp \left ( -\frac{(E-E_{\rm c})^2}
	{E_{\rm width}^2} \right ) \right ]
\exp(-\sigma N_{\rm H}^{\rm circ}) 
\end{equation}
and
\begin{equation}
f_{\rm GIS}(E) = \left [ A \;E^{-\Gamma} + 
   B\exp \left ( -\frac{(E-E_{\rm c})^2}{E_{\rm width}^2} \right ) \right ]
   \exp(-\sigma N_{\rm H}^{\rm circ})
\end{equation}
for the SIS and the GIS data, respectively. Here, $E$ is the X-ray
energy, $f_{\rm Mewe}(kT)$ the model spectrum by Mewe \etal, $\Gamma$
the photon index, $\sigma$ the photo-absorption cross section, $N_{\rm
H}$ the hydrogen equivalent column density, $E_{\rm c}$ the line
center energy, $E_{\rm width}$ the intrinsic line width, which is
assumed to be effectively zero in the fitting, and $A$ and $B$ flux
normalization factors.  The model of the galactic ridge emission,
$f_{\rm Mewe}(kT)$, is included only in the SIS, because it is subtracted
from the GIS spectrum.  The spectra and the best-fit model are shown
in Figure~4, and the best-fit parameters are listed in
Table~3. The luminosity of \gx\ in 2-20 keV (same as the Ginga band) is
estimated as $1.1 \times 10^{37}$ erg s$^{-1}$ using the best-fit
model for the assumed distance of 10 kpc. 

\placefigure{kotani:fig:specasca}

To determine the ionization (state) of iron, we fitted the 5--10 keV
SIS data with a power law attenuated by the iron K edge plus iron
K$\alpha$ and K$\beta$ lines.  The line-flux ratio of the iron
K$\beta$ to K$\alpha$ was fixed to the value expected in fluorescence
of Fe {\sc i}.  The center-energy ratio of the K$\beta$ to K$\alpha$
was also fixed to the value expected in Fe {\sc i}.  We tested several
values of this ratio, and found that they do not change the result.
For SIS0 and SIS1, the center energy of the iron K$\alpha$ line was
fitted to be $6.400^{+0.012}_{-0.010}$ keV and
$6.440^{+0.012}_{-0.012}$ keV, respectively, and the energy of the
iron K edge was fitted to be $7.163^{+0.072}_{-0.072}$ keV and
$7.216^{+0.069}_{-0.072}$ keV, respectively. The errors are 90 \%
confidence intervals.  The line center energies of both sensors were
inconsistent with each other.  This is considered to be due to
gain-calibration uncertainties of the SISs caused by the decreasing
charge transfer efficiency (e.g., Gendreau 1995; Yamashita et al.\
1997).  For measurement of the ionization degree free from the gain
uncertainties, we utilize the ratio of the edge energy to the line
center energy.  The ratios were $1.119^{+0.011}_{-0.011}$ and
$1.120^{+0.011}_{-0.011}$ for SIS0 and SIS1, respectively.  It should
be noted that these values are consistent with each other within the
statistical errors.  Thus the averaged energy ratio was accurately
determined to be $1.120^{+0.008}_{-0.008}$, which corresponds to Fe
{\sc i}-Fe {\sc iv} (Lotz 1968).

\subsubsection{Image analysis}
We searched for a dust scattering halo around \gx.  X-rays from a
source near the galactic plane such as \gx\ would suffer from
scattering by interstellar matter and may produce a halo around the
source.  Efficiency of the dust scattering is a strong function of
X-ray energy and the dust scattering halo is most prominent at lower
energies.  To search for the halo, we compared the radial profile of
the image above and below 5 keV\@.  The dividing energy was selected
to be as low as possible while  retaining statistical significance in
the lower energy band.  This method is unaffected by possible attitude
fluctuation of the satellite during the observations.  Although \gx\ is
located very close to the direction of the galactic center, the effect
of dust scattering was not seen in the image. This is not surprising,
because the source is heavily absorbed and has little X-ray emission
in the lower energy band, where the dust scattering is most prominent.
Predehl, Friedrich, \& Staubert (1995) found an X-ray halo around
\gx in a ROSAT-PSPC image below 2.4 keV\@.  The halo extended to a few
arcmin and was brighter than \gx.  Predehl et al.\ suggested that this 
halo was the reflection of the flare 70 hours before the ROSAT
observation.  That is consistent with our result.

\section{Discussion}
Through the Ginga and ASCA observations, we found large changes in the
energy spectrum, especially for the absorption column density and the
iron emission line.  On the other hand, the spindown rate decreased
slightly in 1990 and 1991 followed by an increased rate, and \gx\
entered a short episode of spinup in 1994 just after the ASCA
observation. The spinup episode is recognized as a glitch in 1995 in
Figure~1 in Chakrabarty et al.\ (1997), or a flare beginning at MJD 49638
in Figure~2 ibid.  In this section, we consider the
possible connection between the circumstellar matter and the spin
period change.

\subsection{Geometry of the circumstellar mater}
Hydrogen column densities obtained from the \gx\ observations with
Ginga and ASCA range between $10^{23}-10^{24}$ cm$^{-2}$.  These
column densities are several orders of magnitude larger than the
galactic absorption of $6\times10^{21}$ cm$^{-2}$ (Dickey \& Lockman
1990), and are considered to have a circumstellar origin. The equivalent
width of the fluorescent iron line is useful to deduce the
distribution of the circumstellar matter.  We show in Figure~5 a
scatter plot of the equivalent width of the iron line and the hydrogen
column density obtained from the ASCA and all the Ginga observations.
The relation expected from a spherical distribution of matter is also
indicated in the figure.  The column density drastically increased to
$10^{24}$ cm$^{-2}$ in 1991 with the increase of iron equivalent width
to 3 keV\@.  This equivalent width is very close to that expected from
a spherical distribution of the matter. The ASCA observation showed
that the column decreased in 1994 to a level comparable to that in
1990.  When the column density is relatively low, the data points are
scattered around the relation expected from the spherical
distribution; but, the deviation is at most factors of a few.  This
scatter may be due to a clumpy structure of the circumstellar
matter. In fact, we found that a partial covering model gives a better
fit to the energy spectrum in 1990.  This supports the assumption that
the circumstellar matter has a clumpy distribution.  Except for this
clumpiness, we can consider that the circumstellar matter has
a spherical symmetric distribution around \gx.

\placefigure{kotani:fig:ew}

We found from the ASCA observation that the ionization degree of iron
in the circumstellar matter is very low (Fe {\sc i}-Fe {\sc iv}).
Thus, the $\xi$-parameter ($\equiv L_X /nr^2$), which determines the
photo-ionization degree, may be smaller than 30 erg cm s$^{-1}$.  With
the knowledge of $L_X$ and $N_H$ ($\sim\!nr$), we can estimate the
characteristic scale of the attenuating-matter distribution or the
typical distance of the matter from the source.  ASCA data are not
very useful to evaluate $L_X$, because the data cover only below 10
keV\@.  According to balloon experiments, the hard X-ray luminosity
(20--100 keV) was approximately $\sim\!2\times10^{37}$ erg s$^{-1}$
(for the distance of 10 kpc) during the 1990's except for the bright state in
1993 December (Rao \etal\ 1994; Paul \etal\ 1997).  Therefore, we
assume $L_X = 2\times10^{37}$ erg s$^{-1}$ as the total X-ray
luminosity during the ASCA observation.  This X-ray luminosity and the
column density of $N_{\rm H} = 2.2 \times 10^{23}$ cm$^{-2}$ give the
radius of the attenuating circumstellar matter to be larger than
$3\times10^{12}$ cm. Thus, it is concluded that the source was
enveloped in a spherical gas  cloud with a radius of $>3\times10^{12}$ cm.
Considering various uncertainties in the above calculation, the result
should be regarded as an order of magnitude estimation.  However, it
is still larger than the co-rotation radius $r_{\rm co} = 4 \times
10^9$ cm and the Alfven radius $r_{\rm A} = 4 \times 10^9$ cm even for
the extraordinarily strong magnetic field of $10^{14}$ G\@.  Instead, it
may be comparable to the size of the Roche lobe, if the binary period
is of order a year.

\subsection{X-ray heating of the circumstellar matter}
\gx\ is thought to accrete matter from the slow stellar wind of the
companion.  The large size of the circumstellar matter and its nearly
spherical distribution around \gx\ is naturally interpreted as the
dense stellar wind from the companion captured in the Roche lobe of
\gx.  
The temperature of the plasma is estimated as a few eV from the value
of the $\xi$-parameter ($\sim\!30$ erg s$^{-1}$ cm), which corresponds
to a thermal velocity of about 30 km s$^{-1}$.  On the other hand,
the Kepler velocity at $10^{12}$ cm from the neutron star is about 100
km s$^{-1}$.  Thus the circumstellar matter is considered to be bound
to \gx.  It is noteworthy that, if \gx\ becomes slightly brighter, the
circumstellar matter can easily escape from \gx.  Suppose \gx\ becomes
3 times more luminous, the $\xi$-parameter of the circumstellar matter
would increase up to $10^2$.  This  corresponds to a plasma temperature of
$\sim\!30$ eV, and hence a thermal velocity of 100 km s$^{-1}$.  This
is comparable to the Keplerian velocity and the plasma can escape from
\gx.

During the spin down phase of \gx\ since 1985, the stellar wind captured
by \gx\ may be stagnated around the neutron star at a distance of
$10^{12}$ cm or more.
The low luminosity characteristic of the spin down phase would reduce
the amount of matter escaping from
\gx\ by  X-ray heating, resulting in a larger column density.
Some other effects such as a change of the wind velocity may also
have contributed to the increase in column density.  Because the amount
of the stellar wind captured by the neutron star depends on the wind
velocity as $\propto v_w^{-4}$, a small change of the wind velocity can
result in the large change of the captured mass.

When \gx\ spins up, as in 1970's and in the short episode of
1994--1995, the X-ray luminosity is at least several times larger than
that in the spindown phase.  The X-ray luminosity may be large enough
to heat up the circumstellar matter and to expel  much of it from the vicinity
of \gx.  This explains the relatively low column density  observed during the
1970's (Becker \etal\ 1976).

Just before the short episode of spinup in 1994-1995, the column
density was observed with ASCA to be reduced to the level of 1990.
This means that most of the accumulated circumstellar matter had been
expelled from the vicinity of \gx\ before the transition to spinup.
Thus it is suspected that the accumulation of the circumstellar matter
may not be the cause of the spinup, and is just controlled by the X-ray
luminosity. This may be understood if the matter which affects
the spin period change is located near the Alfven radius, not at
the distance of $10^{12}$ cm from the neutron star.

\subsection{Optical observations}
Optical observation is powerful for plasma diagnosis, and it is
important to check whether our interpretation of the X-ray data is
consistent with optical data or not.  Here we briefly describe a
preliminary analysis of optical observations, and examine the
consistency between the two bands.  We do not discuss in detail as this will be reported in a forthcoming paper by Greenhill (1998).

V2116 Oph, the optical counterpart of GX 1+4, was observed with the
3.9 meter Anglo Australian Telescope and the RGO spectrograph on 1994 September
25, 11 days after the ASCA observations. The measurements
covered the wavelength range 3700 to 8900 \ang\ with a resolution of 3.2
\ang\@. On September 26 measurements centered on the H$\alpha$ wavelength were
made with a resolution of 1.0 \ang\ FWHM\@. Photometric observations in the
V, H$\alpha$ and I bands were made with the Mount John University
Observatory 1 meter telescope on five nights (August 25 and 27, September
2, and October 3 and 5) covering the time of the ASCA
observations. The H$\alpha$ flux increased slowly during this period and
was about 0.4 magnitudes brighter on October 5. The V--I color changed only
marginally. In view of these relatively small changes we believe our
AAT data are representative of conditions in the source during the
ASCA observations.

The measured and de-reddened fluxes of some prominent emission lines
are listed in Table~4. The reddening correction assumes the same
interstellar extinction as observed by Chakrabarty and Roche
(1997).  The observed H$\alpha$ flux ($1.6\times10^{-13}$ erg cm$^{-2}$
s$^{-1}$) was about half the value observed by Davidsen et al.\ (1977)
during the high state in the mid 1970's and by Chakrabarty \& Roche
(1997) in 1993. The measured H$\alpha$/H$\beta$ line strength ratio of
$\sim$30 was less than 25 \% of the ratio measured by Davidsen et al.\
during the high state. Numerous Balmer, Paschen, He {\sc i}, O {\sc
i}, O {\sc ii}, [O {\sc iii}], Na {\sc i},Ca {\sc ii} and Fe {\sc ii}
lines were seen in emission but higher excitation lines were weak or
undetected. He {\sc ii} 4686 was weak, the C {\sc ii} /N {\sc iii}
feature common in X-ray sources at 4640--4650 \ang\ was not detected; nor
was the 6830 \ang\ Raman line although this was very strong during the
high state and in 1988 (Davidsen et al. 1977; Chakrabarty \& Roche,
1997). The O {\sc i} 8446 line, an indicator of H$\alpha$ optical
depth (Netzer \& Penston 1976) was almost an order of magnitude
fainter relative to H$\alpha$ than during the 1993 measurements by
Chakrabarty \& Roche (1997).

\placetable{kotani:tbl:line}

Following Chakrabarty \& Roche (1997) we
assume that the emission lines are excited by photo-electric
interactions in circumstellar matter by UV emission from the
disk. Further we adopt the hypothesis developed earlier that the
matter is trapped in the gravitational field of the neutron star.  The
H$\alpha$ line width ($\sim$2 \ang\ FWHM after correction for the
spectrograph resolution) is consistent with Doppler broadening due to
matter gravitationally bound to the neutron star at a radius of
$\sim3\times10^{12}$ cm. As noted by Chakrabarty \& Roche (1997) the
presence of Fe {\sc ii} and absence of Fe {\sc iii} lines implies a
temperature in the emission region of $<$30,000 K. The ratio of the
corrected O {\sc i} 8446 to H$\alpha$ emission is $\sim$0.01 indicating
an H$\alpha$ optical depth $\sim$150 (Netzer \& Penston 1976). From
this and the corrected line strength ratios of the first 4 Balmer
lines we find, using the model of Drake \& Ulrich (1980), an electron
density $n_{\rm e} \sim 3\times 10^{10}$ to $10^{11}$ cm$^{-3}$ and a
temperature $\sim$20,000 K in the emission line region. Similar values
for the electron density and plasma temperature are obtained from the
ratios of the He {\sc i} (5876, 6678 and 7065 \ang) emission lines (Proga
et al.\ 1994).  These values are consistent with the hydrogen
equivalent column density $N_{\rm H} = 2\times10^{23}$ cm$^{-2}$ and
plasma temperature derived from the ASCA spectroscopy assuming a
spherical  distribution of matter  with a radius of $\sim 3 \times
10^{12}$ cm about the source.  The densities are one to two orders of
magnitude greater than those estimated by Davidsen et al.\ (1977) and
by Chakrabarty \& Roche (1997) and the radius is an order of
magnitude smaller.  Thus it is shown that the parameters of the  system
derived from the X-ray data, assuming a spherical distribution of circumstellar matter 
coincide with those of the optically emitting plasma.

Since the separation of the neutron star and its giant companion is
expected to be greater than 1 AU ($1.5\times10^{13}$ cm) the matter
in our model can be bound to the neutron star. Clearly, in the
Davidsen et al.\ (1977) model the emission region surrounds the  whole binary
system.

\subsection{A negative feedback mechanism}
The Ginga and ASCA observations suggest that gas  is accumulated in
the vicinity of the source, probably by the stellar wind from the
companion star, and may be blown off by an increase in X-ray luminosity. If
the accreting matter is supplied from the stellar wind as suggested
(e.g., Dotani et al.\ 1989; Davidsen et al.\ 1977), even a small
increase in X-ray luminosity would lead to a decrease in accretion.  This constitutes a negative feedback process and suppresses the
X-ray luminosity. The situation is similar to the self-regulation of
the luminosity at the Eddington limit. \gx\ entered  a low state
in the early 1980's, and, except for occasional short flares (Chakrabarty et al.\
1997), remained faint until the ASCA observation in
1994. During the spindown phase, the activity may be suppressed to
a low level by the negative feedback mechanism. The condition for
the mechanism to work is rather difficult to satisfy compared to
Eddington accretion: (1) The $\xi$-parameter of the accretion flow
needs to be high enough so that the thermal velocity can exceed the
escape velocity. In the case of \gx, the relatively low density of the
accumulated matter, $\gtrsim$ $10^{11}$ cm$^{-3}$ in 1991, resulted in
a large $\xi$-parameter. On the other hand, in binaries where the
accretion matter is supplied by Roche-lobe overflow, the density $\sim
10^{16}$ cm$^{-3}$ is presumably too high for the mechanism to
work. (2) The flow velocity must be lower than the thermal velocity
attainable through X-ray heating. If the ram pressure of the flow is
higher than the thermal pressure,  trapped matter will not be blown off by 
X-ray heating. The typical wind velocity of M giants is 10-30 km
s$^{-1}$ and it is comparable or smaller than the thermal velocity of
a plasma of $\xi \sim 30$.  When \gx\ is in a flare state and $\xi
\gtrsim 100$, the thermal velocity exceeds the wind velocity.
In the case of massive binaries with early type companion stars such
as Cen X-3 or Vela X-1, the wind is supersonic and the negative
feedback  mechanism is not likely to work, even if the sources are wind fed.

Chakrabarty, van Kerkwijk, \& Larkin (1998) reported that the wind
velocity of \gx\ is  as large as $250\pm50$ km s$^{-1}$.
Chakrabarty et al.\ suggested that radiation from the accretion disk
or the neutron star may contribute to the acceleration of the outflow.
That is consistent with the picture of the negative feedback, which
predicts an outflow of circumstellar matter heated by the source.
 However, if the initial velocity of the stellar wind were as large as
250 km s$^{-1}$, our model would fail.

It should be noted that this negative feedback will not keep the X-ray
luminosity constant. The X-ray source blows off only matter
located far from the source with small gravitational binding energy.
Therefore, the fuel supply decreases after a certain time comparable
to the orbital period. Assuming the central mass to be 1.4 M$_\odot$,
the orbital period is estimated to be $10^2$ days at the distance of
$10^{12}$ cm. Such a negative feedback with a time delay would result
in temporal  variability such as oscillations or limit cycles.  This
mechanism predicts that; large X-ray flares  cannot last long because
they even blow off matter close to the source, and that; small flares or
weak activities last longer.  That is qualitatively consistent with the
pulsed flux history reported by Chakrabarty et al.\ (1997).

\subsection{Spindown and spinup phase}
Our interpretation of  the variability in the luminosity and the absorbing
column density through the negative feedback mechanism explains the
very steady spindown trend. The mass supply, averaged over a time
longer than the feedback timescale, would not change very much even if
the mass loss rate of the companion changes considerably.  Either the 
retrograde disk or a propeller effect processes can spindown the neutron star
steadily under the feedback mechanism.

In the spinup phase during the 1970's and in 1994, the feedback mechanism
obviously did not work. In the 1970's, the luminosity was typically as
high as $10^{37}-10^{38}$ erg s$^{-1}$ and was enough to blow off a
stellar wind with a velocity of 10 km s$^{-1}$ and a density of
$<10^{11}$ cm$^{-3}$. The wind could not reach the surface of the
neutron star unless it was as dense as $>10^{14}$ cm$^{-3}$. To avoid 
blow off, the accreting matter had to be supplied by a very dense
stellar wind or by a Roche-lobe overflow. The condition of transition between
spindown and spinup phases is not understood yet. Since the companion
is a low mass star, X-ray irradiation of the surface may affect the
thermal structure and the evolution (e.g., Podsiadlowski 1991;
D'Antona \& Ergma 1993). Harpaz \& Rappaport (1994) show that such a
star may repeat short episodes (1-100 yr) of mass transfer. It is
possible that the spinup and spindown phases of \gx\ are part of the
episodic mass transfers induced by X-ray irradiation.

\section{Conclusion}
GX~1+4 in spindown phase was observed with Ginga in 1990 and 1991, and
with ASCA in 1994, a month before the transition to a short spinup
episode. The behavior of the source in 1990 and 1991 can be summarized
as below:
\begin{itemize}
\item It was found that the absorption column density drastically
increased to $N_{\rm H} = (1.4 \pm 0.2) \times 10^{24}$ cm$^{-2}$ in 1991,
indicating a rapid accumulation of matter in the vicinity of the
source.
\item The luminosity decreased to $L_{\mbox{2-20 keV}} = 0.72 \times 
10^{37}$ erg s$^{-1}$ in 1990, and to $0.27\times 10^{37}$ erg
s$^{-1}$ in 1991, which is almost the same level as the lowest activity in
1987.  Nevertheless, the spindown rate was very stable at $(3-4) \times 
10^{-8}$ s s$^{-1}$ during the time.
\end{itemize}
Properties of the source deduced from the ASCA observation can be
summarized as below:
\begin{itemize}
\item The absorption column density decreased to 
$N_{\rm H} = (2.08 \pm 0.02) \times 10^{23}$ cm$^{-2}$ in 1994 after
a long interval in a high absorption phase.
\item The ionization stage of iron in the absorbing matter was derived 
to be Fe {\sc i}--Fe {\sc iv} from the ratio of the absorption-edge
energy to the emission-line energy.
\item  The $\xi$-parameter of the absorbing matter was
\raisebox{-.5ex}{$\lesssim$}30, consistent with a spherical
distribution with a radius of $>3\times 10^{12}$ cm about the source.
\item The luminosity brightened to $L_{\mbox{2-20 keV}} = 1.1
\times  10^{37}$ erg s$^{-1}$, which is the brightest in the spindown 
phase, and almost comparable to that in the highly-active spinup
phase.
\end{itemize}
The picture of \gx\ in a gas cloud is also supported by the optical
observations covering the ASCA observation.  The Doppler broadening of
the H$\alpha$ line was consistent with the Kepler motion at
$\sim3\times 10^{12}$ cm from a neutron star, and electron density and
temperature estimated from the line strength ratios were consistent
with the values expected from the picture.

To explain the accumulation and the disappearance of the absorbing
matter, a negative feedback mechanism stabilizing the spindown rate is
suggested.  Gas supplied by a slow stellar wind from the companion
star would be blown off, if the luminosity of the neutron star is
high.  When the luminosity is low, the neutron star would accrete and
increase the luminosity.  Therefore, the accretion rate and thus the
spindown rate may be stabilized under a certain level.  This mechanism
does not depend on the nature of the unknown spindown mechanism, for
which a retrograde disk and high magnetic field were proposed.

Since this negative feedback mechanism will not explain transitions
between spinup and spindown, the transitions should be explained by
some other mechanism such as episodic mass transfers.

\vspace{1cm}

TK would like to thank Dr.\ T. Aoki, Dr.\ K. Asai and Dr.\ H.
Negoro, for helping to analyze Ginga data. TK is supported by the
Special Postdoctoral Researchers Program of RIKEN\@.

\clearpage
\noindent
{\large\bf References}
\begin{description}

\item Becker, R. H., Boldt, E. A., Holt, S. S., Pravdo, S. H.,
	Rothschild, R. E., Serlemitsos, P. J., \& Swank, J. H. 1976, 
	ApJ, 207, L167

\item Burke B. E., Mountain R. W., Daniels P. J., Cooper M. J., \&
	Dolat V. S. 1994, IEEE Trans.\ Nucl.\ Sci.\ 41 (1), 375

\item Chakrabarty D., Prince T. A., Finger M. H., \& Wilson R. B. 1994,
	IAUC No.\ 6105

\item Chakrabarty D., Koh T., Prince T. A., Vaughan B., Finger M. H., 
	Scott M., \& Wilson R. B. 1995, IAUC No.\ 6153

\item Charkrabarty, D., Bildsten L., Finger, M. H., Grunsfeld, J. M.,
	Koh, D. T., Nelson, R. W., Prince, T. A., Vaughan, B. A., \& 
	Wilson, R. B. 1997, ApJ 481, L101

\item Chakrabarty D.  \& Roche P. 1997, ApJ 489, 254

\item Chakrabarty D., van Kerkwijk M. H., \& Larkin J. E. 1998, ApJ 497, L39

\item Cutler E. P., Dennis B. R.,  \& Dolan J. F. 1986, ApJ 300, 551

\item D'Antona F.  \& Ergma E., 1993, A\&A 269, 219

\item Davidsen A., Malina R.,  \& Bower S. 1977, ApJ 211, 866

\item Dotani T., Kii T., Nagase F., Makishima K., Ohashi T., Sakao T.,
	Koyama K.,  \& Tuohy I. R. 1989, PASJ 41, 427

\item Dickey J.M. \&  Lockman F.J. 1990, Annual Review of Astron.\ 
Astrophys. 28, 215

\item Drake S. A. \&  Ulrich R.K. 1980, ApJS 42, 351

\item Elsner R. F., Weisskopf M. C., Apparao K. M. V., Darbro W.,
	Ramsey B. D., \&  Williams A. C. 1985, ApJ 297, 288

\item Finger M. H., Wilson R. B., Fishman G. J., Bildsten L., Chakrabarty D., \& 
	Prince T. A. 1993, IAUC No.\ 5859

\item Gendreau, K. C. 1995, Ph.D. Thesis, MIT

\item Ghosh P. \& Lamb F. K., 1979a, ApJ 232, 259

\item Ghosh P. \& Lamb F. K., 1979b, ApJ 234, 696

\item Goldwurm A., Denis M., Paul J., Faisse S., Roques J. P.,
	Bouchet L., Vedrenne G., Mandrou P., \&  Sunyaev R., \etal\
	1995, Adv.\ Space Res.\ 15 (5), 41

\item Greenhill J. G., Sharma D. P., Dieters S. W. B., Sood R. K., 
	Waldron L., \&  Storey M. C. 1993, MNRAS 260, 21

\item Greenhill J. G. 1998, MNRAS, in preparation

\item Harpaz A. \&  Rappaport S., 1994, ApJ 434, 283

\item Hayashida K., Inoue H., Koyama K., Awaki H., Takano S., Tawara Y., 
	Williams O. R., Denby M., Stewart G. C., Turner M. J. L.,
	Makishima K., \&  Ohashi T. 1989, PASJ 41, 373

\item Koyama, K. 1989, PASJ 41, 665

\item Lotz, W. 1968, J. Opt. Soc. Amer. 58, 915

\item Lutovinov A., Grebenev, S., Sunyaev R., \&  Pavlinsky M. 1995,
	Adv.\ Space Res.\ 16 (3), 135

\item Makino F., \&  the ASTRO-C Team 1987, Astrophys.\ Letters Commun.\ 25, 223

\item Makishima K. 1986, in The Physics of Accretion onto Compact Objects,
	ed.\ K.O. Mason, M.G. Watson, \& N.E. White (Berlin: Springer), 249

\item Makishima K., Ohashi T., Sakao T., Dotani T., Inoue H., Koyama K., 
	Makino F., Mitsuda K., et al. 1988, Nature 333, 746

\item Makishima K., Tashiro M., Ebisawa K., Ezawa H., Fukazawa Y.,
	Gunji S., Hirayama M., Idesawa E., et al.\ 
	1996, PASJ 48, 171

\item Mandrou P., Roques J. P., Bouchet L., Niel M., Paul J., Leray J. P.,
	Lebrun F., Ballet J., \etal\ 1994, ApJS 92, 343

\item McClintock J. E., \& Leventhal M. 1989, ApJ 346, 143

\item Mewe R.,   Gronenschild E. H. B. M., van den Oord G. H. J. 1985,
 A\&AS, 62, 197

\item Mony B., Kendziorra E., Maisack M., Staubert R., Englhauser J.,
	D\"{o}bereiner S., Pietsch W., \& Reppin C. \etal\ 1991, A\&A
      257, 405

\item  Nelson R. W., Bildsten L., Chakrabarty D., Finger M. H., Koh
D. T., Prince T. A., Rubin B. C., Scott D. Mathew., Vaughan B. A., \&
Wilson R. B. 1997, ApJL 488, L117

\item Netzer H., \& Penston M. V. 1976, MNRAS 174, 319

\item Paul, B., Agrawal, P. C., Rao, A. R., \& Manchanda, R. K. 1997, A\&A
	319, 507

\item Podsiadlowski Ph., 1991, Nature 350, 136

\item Predehl, P., Friedrich, S., \& Staubert, S. 1995, A\&A 294, L33

\item Proga D., Mikolajewska J., \& Kenyon S. J. 1994, MNRAS 268, 213

\item Rao, A. R., Paul, B., Chitnis, V. R., Agrawal, P. C.,  \&
	Manchanda, R. K. 1994, A\&A 289, L43

\item Sakao T., Kohmura Y., Makishima K., Ohashi T., Dotani T., 
	Kii T., Makino F., Nagase F., et al.
	1990, MNRAS 246, 11p

\item Serlemitsos P. J., Jalota L., Soong Y., Kunieda H., Tawara Y.,
	Tsusaka Y., Suzuki H., Sakima Y., et al.\ 1995,
	PASJ 47, 105

\item Sood R. K., James S. D., Lawson W. A., Sharma D. P., Grey D. G., \&
	Manchanda R. K. 1995, Adv.\ Space Res.\ 16 (3), 131

\item Staubert R., Maisack M., Kendziorra E., Draxler T., Finger M. H., 
	Fishman G. J., Strickman M. S., \& Starr, C. H. 1995,
	Adv.\ Space Res., 15 (5), 119

\item Tanaka Y., Inoue H., \& Holt, S. S. 1994, PASJ, 46, L37

\item Turner M. J. L., Thomas H. D., Patchett B. E., Reading D. H., 
	Makishima K., Ohashi T., Dotani T., Hayashida K., et al.\ 1989, 
	PASJ 41, 345

\item Wang Y. -M. 1987, A\&A 183, 257

\item Yamashita A., Dotani T., Bautz M., Crew G., Ezuka H., Gendreau K.,
Kotani T., Mitsuda K., Otani C., Rasmussen A., Ricker G., Tsunemi H.
1997, IEEE Transactions Nucl.\ Sc.\ 44, 847

\item Yi, I., Wheeler, J. G., \& Vishniac, E. T. 1997, ApJ 481, L51
\end{description}

\clearpage

\figcaption[f1a.eps,f1b.eps] {Folded light curves of the Ginga data in 
(a) 1990 September (1--37 keV) and (b) 1991 September (2--60 keV)\@.
Background and galactic-ridge components were not subtracted.
See Table~2 for the heliocentric pulse period of these data.
\label{kotani:fig:lc}}

\figcaption [f2a.eps,f2b.eps] {Energy spectra taken with Ginga in 
(a) 1990 September and (b) 1991 September. The intrinsic and the
galactic background are subtracted. The best-fit model (solid) and the
contribution of the power law (dot) and the iron line (dot-dash) are
plotted as histograms.  The contribution of leaked power law (dash) is
also plotted in (a).  See Table~3 for the best-fit
parameters. \label{kotani:fig:specginga}}

\figcaption [f3.eps] {Folded light curves of the SIS data in six
energy bands in 1994 September. The temporal average of the flux of
each curve is normalized to unity.  Both sensors are combined. The
epoch at phase 0 is 49610.00 (MJD). Background and galactic-ridge
components were not subtracted. \label{kotani:fig:lcasca}}

\figcaption [f4a.eps,f4b.eps] {Spectra taken with ASCA (a) SIS and (b) 
GIS. The best fit model is plotted as histogram. See Table~3 for the
best fit parameters. From the GIS data, the galactic ridge emission is
subtracted, while only the diffuse cosmic X-ray background component is
subtracted from the SIS data.
\label{kotani:fig:specasca}}

\figcaption [f5.eps] {History of the equivalent width and the
absorption column density. The data before 1989 were taken from Sakao
et al.\ (1990). The column density of 1990 was multiplied by the
covering fraction. The relation of equivalent width and absorption
column density expected in the case of isotropic distribution of
absorbing matter (Makishima 1986) is plotted as a dashed line.
\label{kotani:fig:ew}}

\clearpage
\begin{table}
\caption{Journal of observations}\label{kotani:tbl:obslog}
\begin{center}
\begin{tabular}{clcccc} \hline\hline
No & Satellite & Start Time & End Time & Exposure & Energy Range\\
   &           & (UT)       &(UT)      & (ks)       & (keV)\\
 \hline
1 & Ginga & 1990/9/4 13:47 & 1990/9/5 04:28 & 10.8 & 1--37 \\
2 & Ginga & 1991/9/14 16:56 & 1991/9/16 01:41 & 37.8 & 2--60 \\ 
3 & ASCA  & 1994/9/14 23:53 & 1994/9/15 22:00 & 30.0 & 0.5--10 \\ \hline\hline
\end{tabular}
\end{center}
\end{table}

\clearpage
\begin{table}
\caption{Results of timing analysis}\label{kotani:tbl:tim}
\begin{center}
\begin{tabular}{lcccc} \hline\hline
No & Mission & Epoch           & Pulse period & Period derivative\dag \\
   && (MJD)     & (s)              & ($10^{-8}$ s s$^{-1}$) \\ \hline
1  &Ginga& 48138.0   & $114.63 \pm 0.02$\ddag  & $3.1 \pm 0.1$ \\
2  &Ginga& 48513.0   & $116.17 \pm 0.03$\ddag  & $4.7 \pm 0.2$ \\
3  &ASCA& 49610.5   & $122.02\pm0.03$\S    & $6.17 \pm 0.03$\\
 \hline\hline
\multicolumn{4}{l}{\dag Average between observations.}\\
\multicolumn{4}{l}{\ddag Heliocentric value.}\\
\multicolumn{4}{l}{\S Barycentric value.}\\
\end{tabular}
\end{center}
\end{table}

\clearpage
\begin{table}
\caption{Spectral parameters of \gx}\label{kotani:tbl:best}
\begin{center}
\begin{tabular}{lccccc} \hline\hline
No& \multicolumn{2}{c}{Power Law}           & Absorption          & \multicolumn{2}{c}{Iron Line} \\
  & Normalization\P      & Photon index     & N$\rm_H$            & EW                  & Center \\
  &                      &                  &($10^{23}$ cm$^{-2}$)& (keV)               & (keV) \\ \hline
1 &$0.156\pm0.026$       &$1.98\pm0.07$     &$3.0\pm0.3$          &$0.11\pm 0.03$	&6.4 (fixed)\\
  &                      &                  &$A$\dag=$0.87$       &	                &\\
2 &$0.022\pm0.015$       &$1.49\pm0.23$     &$14\pm2$	          &$3.1\pm 0.2$         & 6.4 (fixed)\\
3 &$0.0503\pm0.0026$\ddag&$1.19\pm0.09$\ddag&$2.15\pm 0.03$\ddag  &$0.094\pm0.005$\ddag	&$6.42\pm0.04$\S\\
\hline\hline
\multicolumn{6}{l}{\P In unit of photons cm$^{-2}$ s$^{-1}$ keV$^{-1}$ at 1 keV}\\
\multicolumn{6}{l}{\dag Partial covering factor.}\\
\multicolumn{6}{l}{\ddag Average between SIS and GIS.}\\
\multicolumn{6}{l}{\S SIS only. Including systematic errors.}
\end{tabular}
\end{center}
\end{table}

\clearpage
\begin{table}
\caption{Prominent emission lines}\label{kotani:tbl:line}
\begin{center}
\begin{tabular}{cccc}\hline\hline
Ion	&Wavelength &Measured Flux	&Corrected Flux\dag\\
	&(\ang)	&($10^{-16}$ erg cm$^{-2}$ s$^{-1}$)	&($10^{-14}$ erg cm$^{-2}$ s$^{-1}$)\\
\hline
H {\sc i} &6563	&1590	&619\\
H {\sc i} &4861	&52	&122\\
H {\sc i} &4341	&6.6	&31.4\\
H {\sc i} &4102	&2.6	&17.2\\
He {\sc i} &5876	&39	&28.3\\
He {\sc i} &6675	&27	&9.6\\
He {\sc i} &7065	&54	&14.4\\
He {\sc ii} &4686	&3.2	&9.5\\
O {\sc i} &8446	&50	&6.0\\
\hline
\multicolumn{4}{l}{Measured at the AAT on 1994 September 26.}\\
\multicolumn{4}{l}{\dag The reddening corrections
are those used by Chakrabarty and Roche 1997}
\end{tabular}
\end{center}
\end{table}

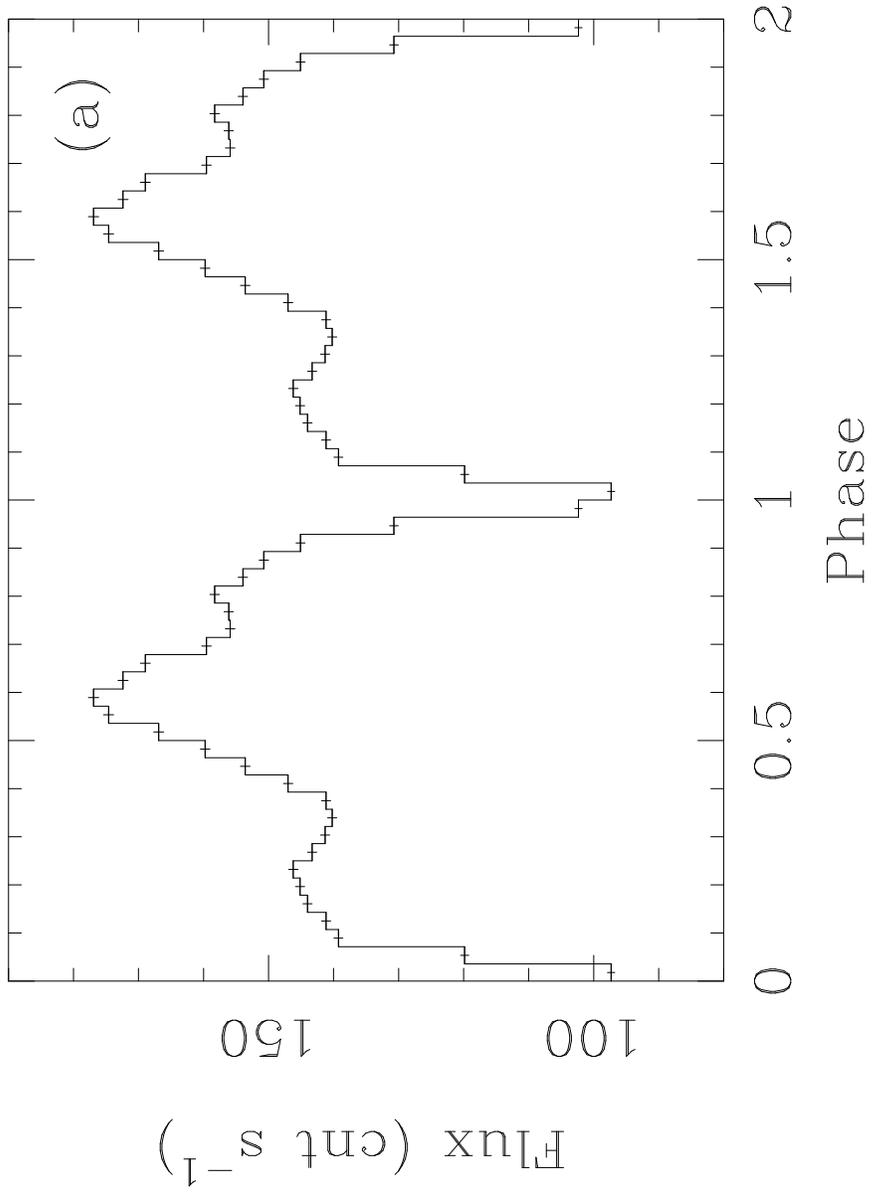
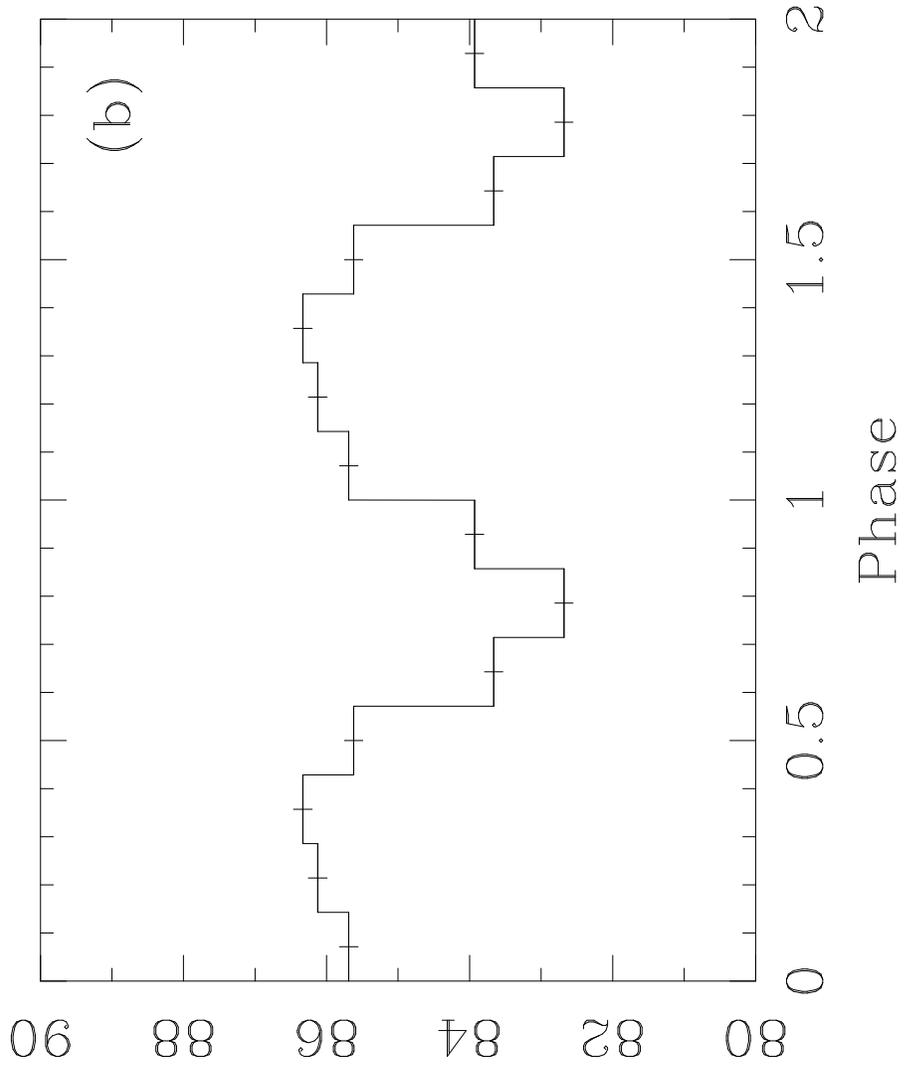
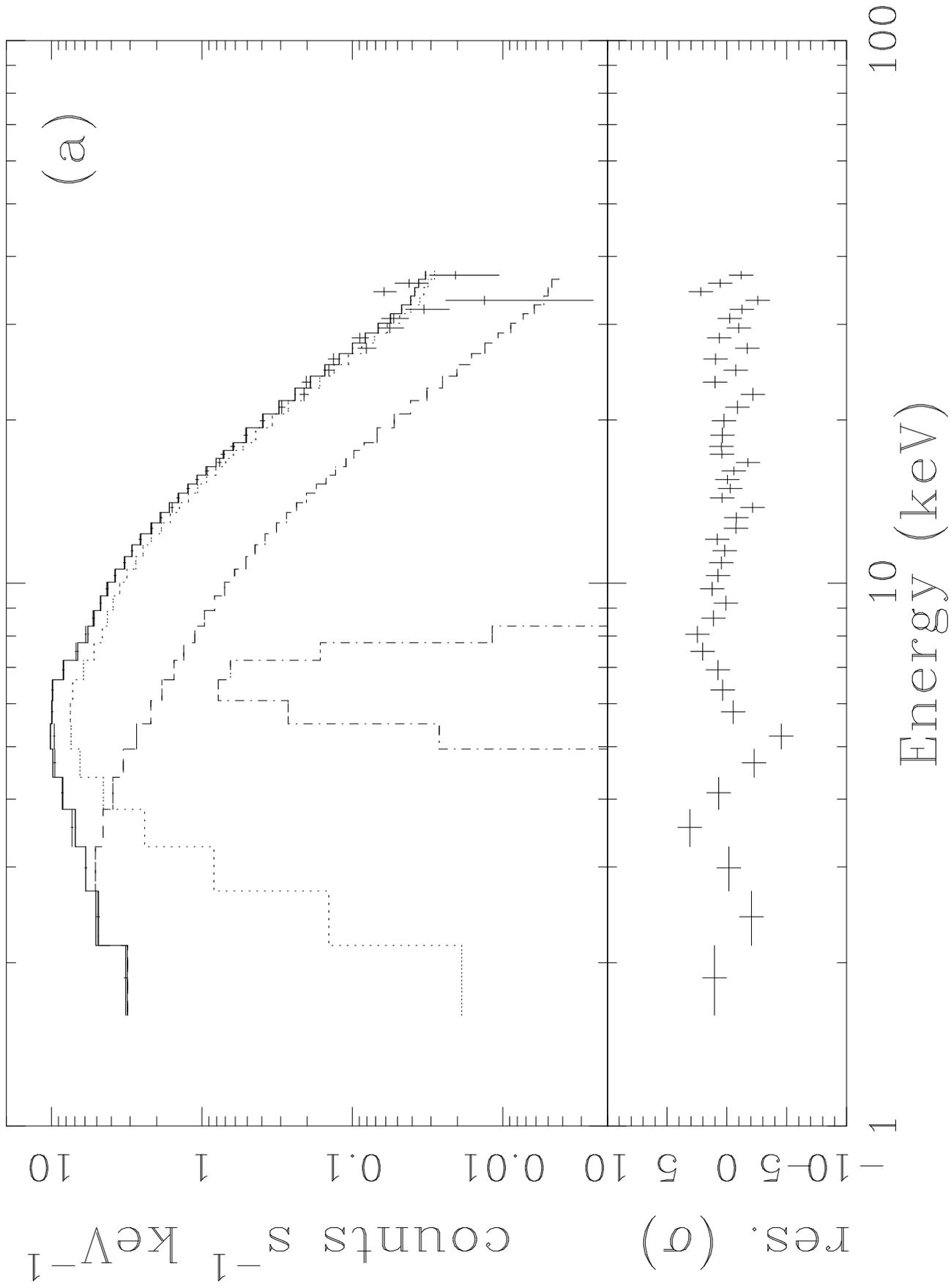
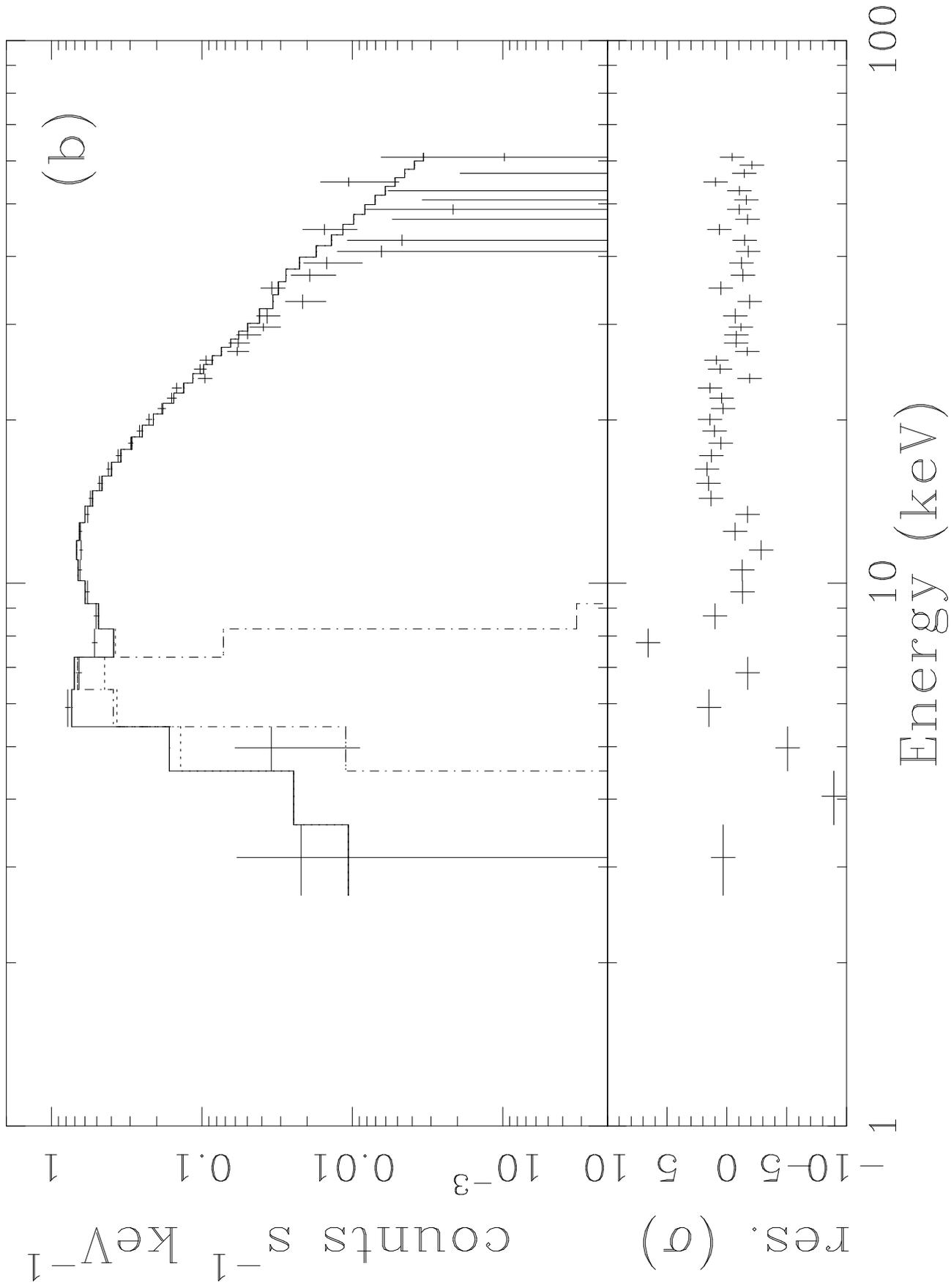
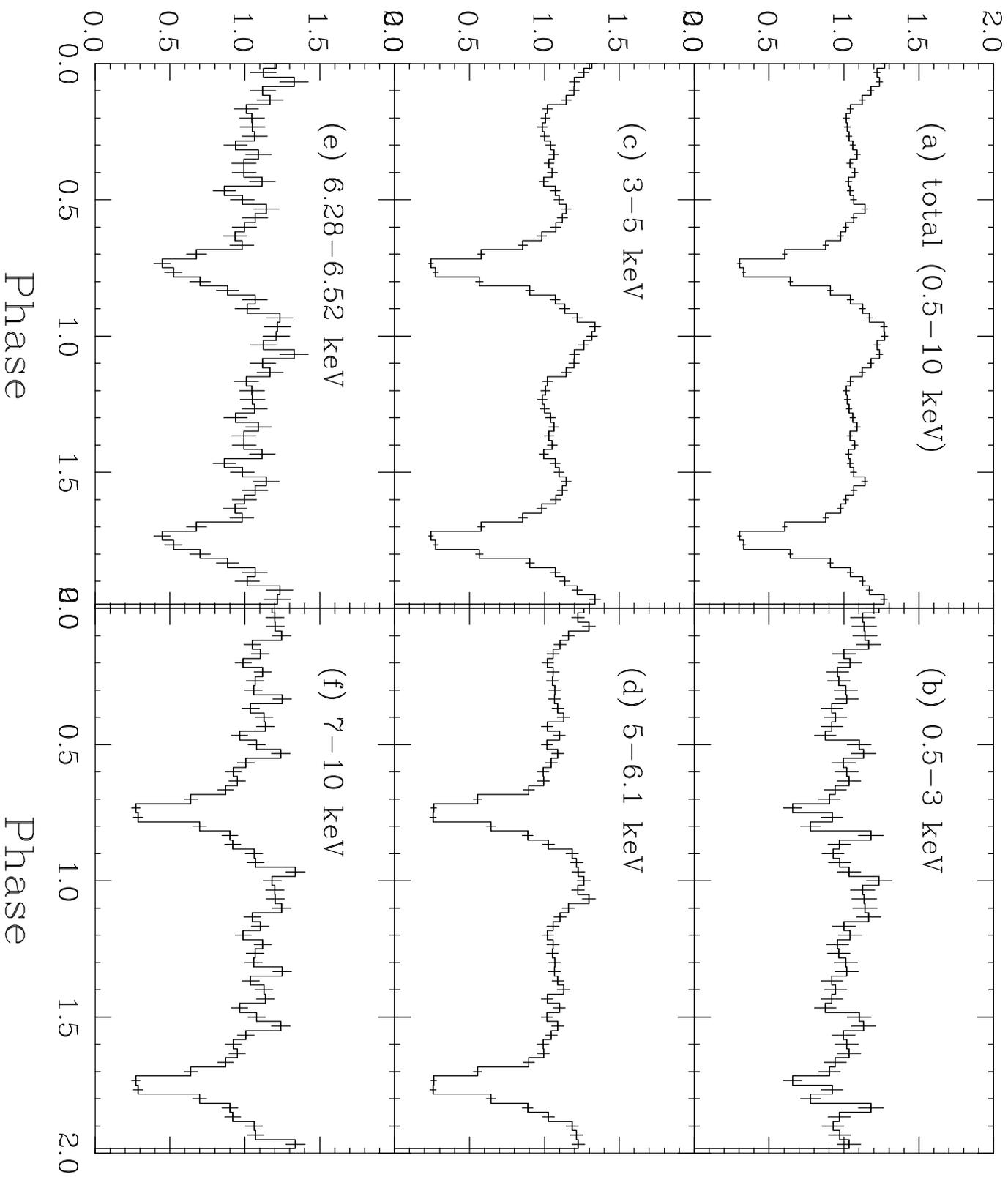
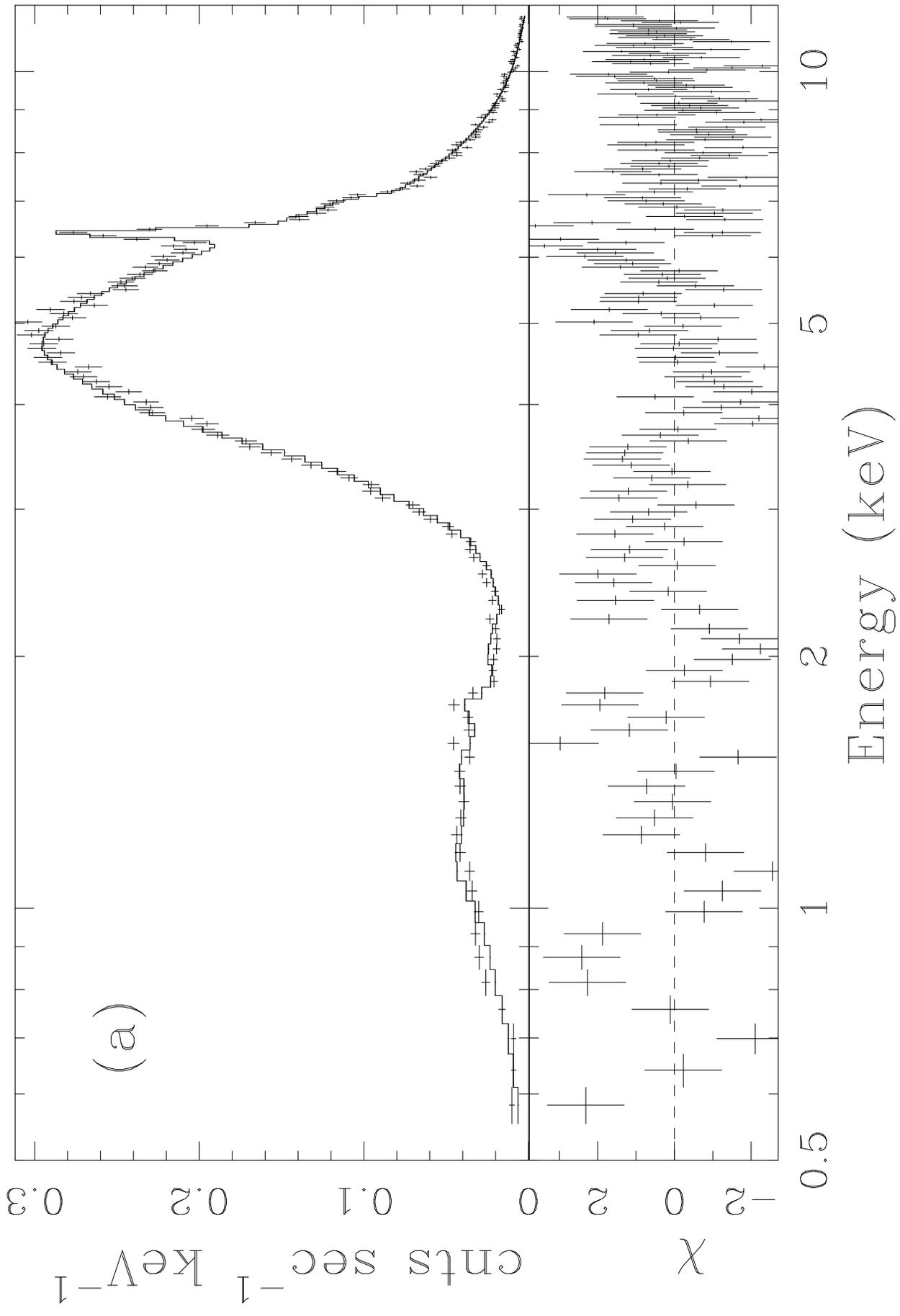
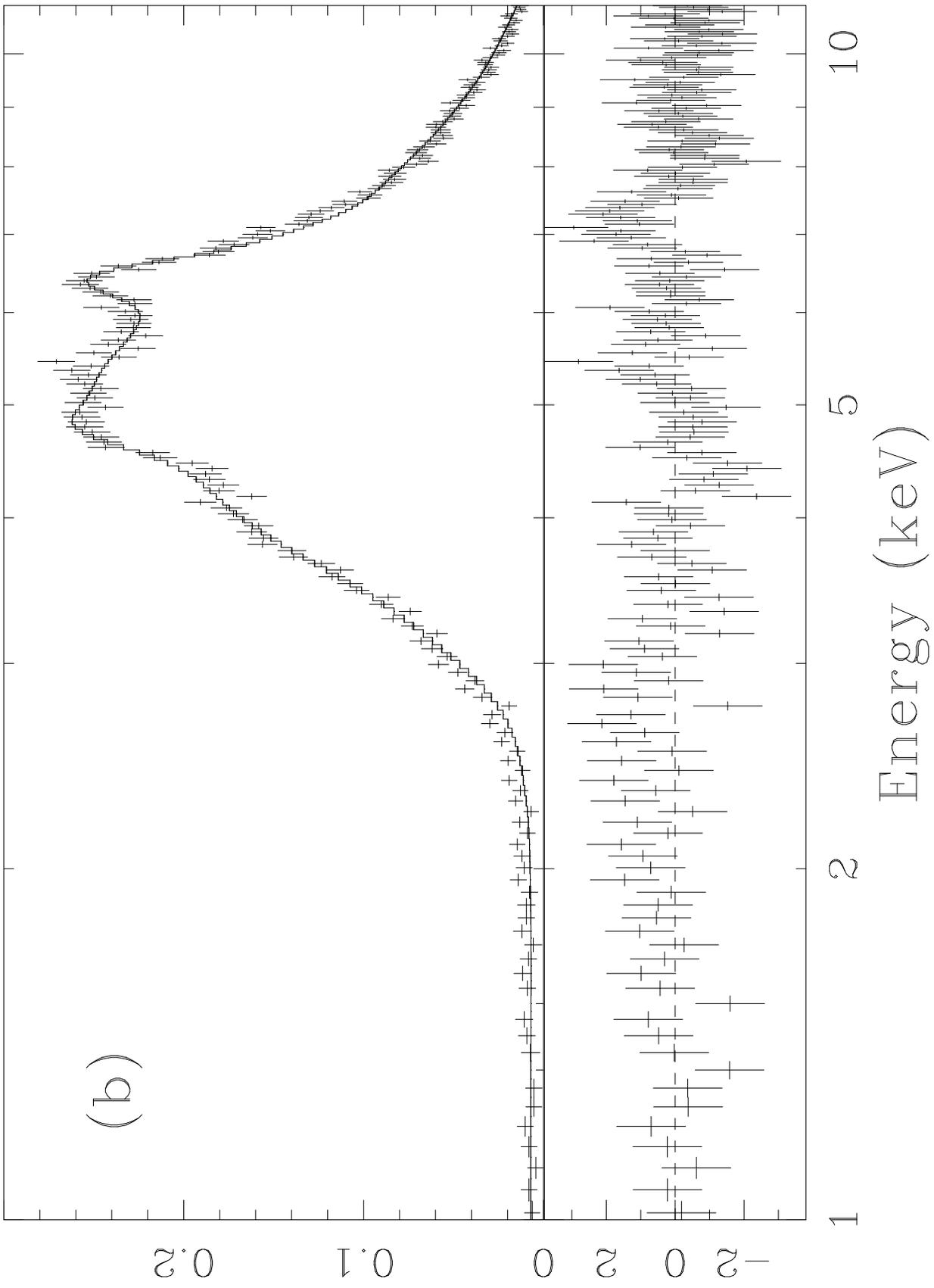
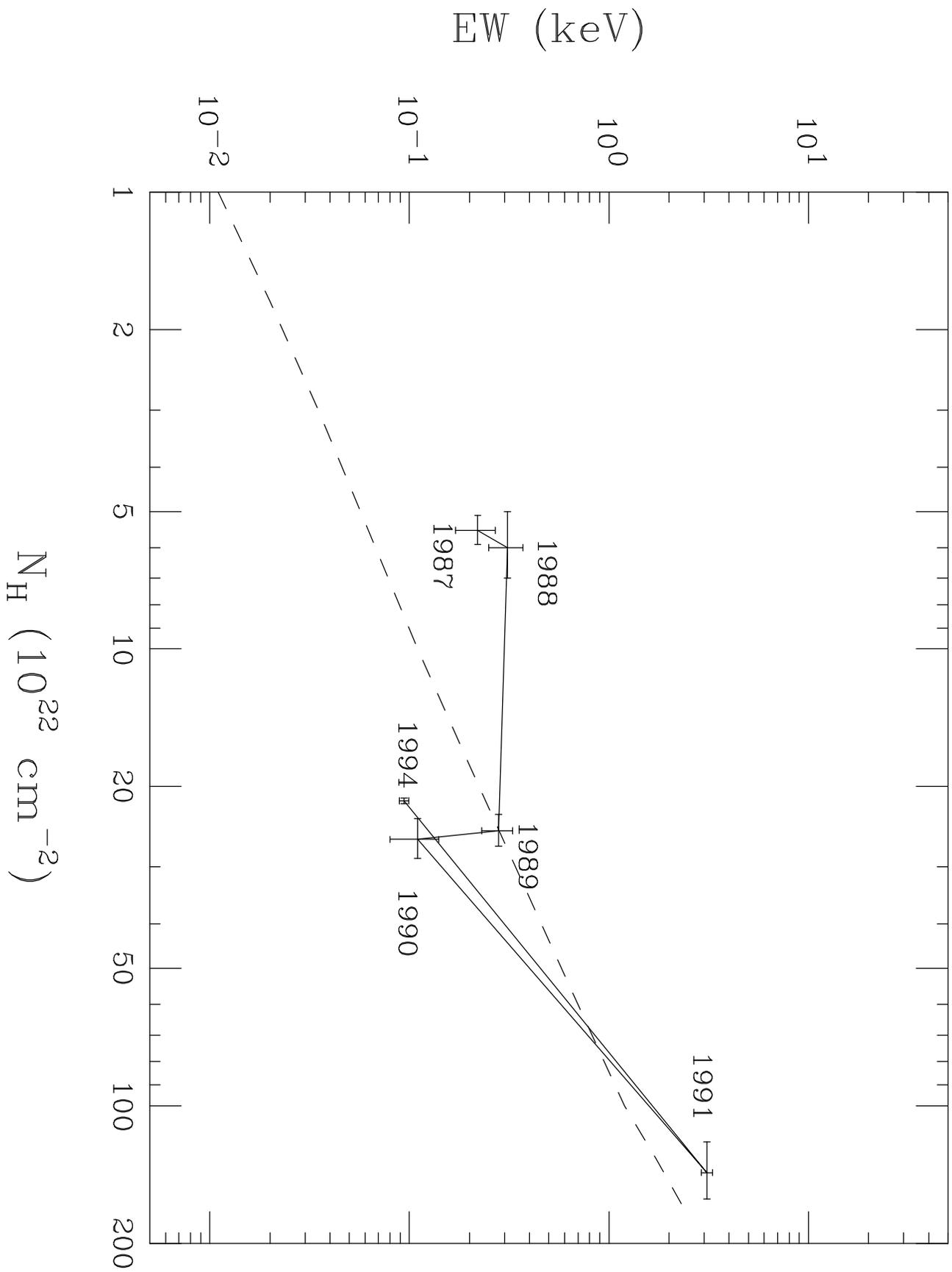
\end{document}